\documentclass[twocolumn,journal]{IEEEtran}
\IEEEoverridecommandlockouts

\usepackage{cite}
\usepackage{amsmath,amssymb,amsfonts}
\usepackage{graphicx}
\usepackage{textcomp}
\usepackage{xcolor}
\usepackage{bm,bbm}
\usepackage{algorithm}
\usepackage{algpseudocode}
\usepackage{url}
\usepackage{epstopdf} 

\usepackage{footnote}
\usepackage{setspace}
\usepackage{comment}
\usepackage{soul}
\allowdisplaybreaks

\algrenewcommand\algorithmicrequire{\textbf{Input:}}
\algrenewcommand\algorithmicensure{\textbf{Initialize:}}
\newtheorem{assumption}{Assumption}

\newtheorem{theorem}{Theorem}
\newtheorem{example}{Example}

\def\BibTeX{{\rm B\kern-.05em{\sc i\kern-.025em b}\kern-.08em
    T\kern-.1667em\lower.7ex\hbox{E}\kern-.125emX}}

\begin{document}

\title{Ordering for Communication-Efficient Quickest Change Detection in a Decomposable Graphical Model\\
\thanks{The work is supported by the U. S. Army Research Laboratory and the U. S. Army Research Office under grant number
W911NF-17-1-0331, the National Science
Foundation under Grant ECCS-1744129, and a grant from the Commonwealth of Pennsylvania, Department of Community and Economic Development, through the Pennsylvania Infrastructure Technology Alliance (PITA). {Some preliminary work was presented in Y. Chen, R. Blum, and B. Sadler, ``Optimal quickest change detection in sensor networks using ordered transmissions,'' IEEE Workshop on Signal Processing Advances in Wireless Communications, 2020.}}}

\author{Yicheng Chen, Rick S. Blum,
\IEEEmembership{Fellow,~IEEE}, and Brian M. Sadler, \IEEEmembership{Fellow,~IEEE}
\thanks{
Yicheng Chen and Rick S. Blum are with Lehigh University, Bethlehem, PA 18015 USA
(email: yic917@lehigh.edu, rblum@eecs.lehigh.edu).}
\thanks{
Brian M. Sadler is with the Army Research Laboratory, Adelphi, MD 20878 USA (email: brian.m.sadler6.civ@mail.mil).}}


\maketitle

\begin{abstract}
A quickest change detection {problem} is considered in a sensor network with observations whose statistical dependency structure across the sensors before and after the change is described by a decomposable graphical model (DGM). Distributed computation methods for {this problem} are proposed {that} are capable of producing the optimum centralized test statistic. {The DGM leads to} the proper way to collect nodes into local groups equivalent to cliques in the graph, such that a clique statistic which summarizes all the clique sensor data can be computed within each clique. The clique statistics are transmitted to a decision maker {to produce the optimum centralized test statistic}. In order to further improve communication efficiency, an ordered transmission approach is proposed where transmissions of the clique statistics {to the fusion center} are ordered and then adaptively halted when sufficient information is accumulated.  {This procedure is always guaranteed to provide the optimal change detection performance, despite not transmitting all the statistics from all the cliques.} A lower bound on the average number of transmissions saved by ordered transmissions is provided and for the case where the change seldom occurs the lower bound approaches approximately half the number of cliques provided a well behaved distance measure between {the distributions of the sensor observations} before and after the change is sufficiently large. {We also extend the approach to the case when the graph structure is different under each hypothesis.} Numerical results show significant savings using the ordered transmission approach and validate the theoretical findings.
\end{abstract}


\IEEEpeerreviewmaketitle

\section{Introduction}
Sensor networks are critical for many applications such as disaster {response}, security, smart cities, enhanced
building operation for optimized energy usage, {health monitoring and assisted living},
and smart transportation systems \cite{akyildiz2002wireless, chong2003sensor}. A fundamental problem is to detect the occurrence
of a change. This can be modeled as
a quickest change detection (QCD) problem, see \cite{veeravalli2014quickest,mei2010efficient,banerjee2015data,lorden1971procedures,lai1998information,pollak1985optimal,veeravalli2001decentralized,mei2005information} and references therein.

The classical {centralized and unconstrained communication} QCD problem in sensor networks is well investigated \cite{veeravalli2014quickest,veeravalli2001decentralized,mei2005information,tartakovsky2008asymptotically} where each sensor monitoring the environment takes a sequence of observations. At an unknown change time, the distribution of the observations at all sensors change simultaneously. Based on the data received from the sensor nodes, a decision maker at a fusion center (FC) would like to detect the change as soon as possible subject to {a} false alarm {constraint}. Depending on knowledge of the change time distribution, Bayesian and minimax QCD formulations have been developed, and the corresponding optimal solutions are introduced in \cite{veeravalli2014quickest}. In this paper, we focus on the minimax formulation where we model the change time as a deterministic but unknown positive integer and minimize the worst case average detection delay (WADD) subject to a false alarm constraint. {In many cases}, each sensor in the network carries its own limited energy source, {and the energy cost of communications is significant}. Hence, communication efficiency is an important topic for QCD in sensor networks.

A particularly popular approach called censoring has been shown to be an effective method to improve communication efficiency where sensors transmit only highly informative data \cite{rago1996censoring}. In \cite{rago1996censoring}, {upper and lower thresholds are set and} sensors transmit only very large or small likelihood ratios {because these values provide} significant information about which hypothesis is most likely to be true.  Censoring-based QCD is proposed in \cite{banerjee2015data} \cite{mei2011quickest} {where it is shown that} censoring yields transmission savings but always increases detection delay {(the accepted performance measure {for QCD})}.

{In this paper we introduce an ordered transmission QCD method that will lower communications without increasing the detection delay.} {The ordered transmission approach} (also called ordering) was first introduced for a {distributed} testing problem between two fixed hypotheses and {employing an FC} \cite{blum2008energy}. {Using ordering} the sensors with the most informative observations transmit
first. Transmissions can be halted when sufficient information is accumulated for the FC to decide which hypothesis is true. In \cite{blum2008energy}, it was shown that this ordered transmission approach can reduce the number of transmissions without losing any detection performance for cases with statistically independent observations. The detection of a mean shift or covariance matrix change in statistically dependent Gaussian observations following a decomposable Gaussian graphical model (GGM) is considered in \cite{zhang2017ordering} and \cite{chen2019testing}, respectively. In this paper, we provide the first {communication} efficient QCD algorithm for rapidly detecting a change with statistically dependent observations across the sensors. The focus is on the case where the observations follow a decomposable graphical model (DGM) to characterize the dependence among sensor observations, {completely subsuming} the case with independent observations. This work is a highly nontrivial extension of the initial work in \cite{Chen2005:Optimal}, {which was limited to independent observations and {one-hop} communications to an FC}.

\subsection{Our Contributions}

This work describes a new algorithm that provides a
communication efficient distributed processing method for a sensor networking change detection problem where the
observations follow a statistically dependent distribution before and after
the change.  The focus is on the case where both distributions are characterized by a decomposable graphical model (DGM) {with common graph
structure}. {In Section \ref{graphchangescase}, extensions are described for the case where the graph structure of the distributions before and after the change are different}. The distributed computation describes the proper way to collect nodes into local groups, corresponding to the cliques\footnote{{A clique {is defined in this paper} as a set of vertices that induce the largest
fully connected subgraph.}} in the graph, such that each clique can produce a compressed clique statistic that summarizes all the clique sensor data.  The clique test statistics can be transmitted to a common location and summed to produce the standard optimum centralized change detection test statistic. Distributed computation methods have not been previously described for these change detection problems as they are complicated by the statistically dependent data.
{With proper clique formation, we then employ ordering} to reduce the number of clique test statistics that need to be transmitted without any performance loss.  {We derive a lower bound on the number of transmissions saved, and show that up to one half savings is possible for some cases of interest.}

\subsection{Problem Formulation}
We consider a sensor network with $M$ sensors and a FC. Sensor $m$ for $m=1,2,..,M$ observes the sequence $\{X_{n,m}\}_{n\ge1}$ with $n$ {denoting} the time slot index. The objective is to detect a change as quickly as possible after the change occurs which implies the goal is to minimize detection delay if the change occurs. As long as no change is declared, the sensors will continue observing data. Throughout this paper we make the following assumption.
\begin{assumption}\label{assumption3}
The distributions of the observations at all sensors change simultaneously at the change time $\tau$. In particular, at the unknown time slot $\tau$, the distribution of ${X_{n,[1,M]}} \buildrel \Delta \over = {( {{X_{n,1}},{X_{n,2}},...,{X_{n,M}}} )}$ changes from $f_0$ to $f_1$ where $f_0$ and  $f_1$ are the known probability density functions (pdfs) before and after the change time, respectively. {The random variable $X_{n,m}$ is independent across the time slot index $n$ {but will generally assumed to be} dependent across the sensor index $m$.}
\end{assumption}

Without a prior on the distribution of the change time, we model the change time $\tau$ as a deterministic but unknown integer and we employ the {constraint}
\begin{align}\label{falsealarmrate}
\mathbb{E}_{\infty}({n'})\ge\gamma
\end{align}
where $n'$ denotes the time slot when the decision maker declares a change has occurred, $\mathbb{E}_{\infty}(n')$ is the average delay when the change does not occur, and $\gamma$ is a pre-specified constant. If the change occurs, then one formulation to evaluate the detection delay is to employ the WADD defined in \cite{lorden1971procedures} as
\begin{align}\label{WADD}
\mbox{WADD}({n'})=\sup _{\tau \geq 1} \text { ess } \sup \mathbb{E}_{\tau}\left[({n'}-\tau)^{+} | \mathcal{I}_{\tau-1}\right]
\end{align}
where $\text{ess}\sup X$ denotes essential supremum of $X$, $\mathbb{E}_{\tau}$ is the expectation when the change occurs at time $\tau$, $(x)^+ \buildrel\Delta\over=\max\{x,0\}$, $\mathcal{I}_{\tau-1}\buildrel\Delta\over=(X_{[1,\tau-1],1},...,X_{[1,\tau-1],M})$ denotes past global information at time slot $\tau$, and $X_{[1,\tau-1],m}\buildrel\Delta\over=(X_{1,m},...,X_{\tau-1,m})$ denotes past local information at sensor $m$. Thus, the QCD problem in a minimax setting can be formulated as a constrained optimization problem
\begin{align}\label{CADD1}
&\min_n \mbox{WADD}({n'})\nonumber\\
&s.t.\  \mathbb{E}_{\infty}({n'})\ge\gamma.
\end{align}

\subsection{Paper Organization}
The paper is organized as follows. In Section \ref{DGMsection}, a brief discussion on mathematical formulations of {decomposable graph models} is described. In Section \ref{LIkelihood}, we {describe distributed computation of the optimum test statistic}. {Communication}-efficient QCD using ordered transmissions is described in Section \ref{EnergyEfficientQCDOrdering} {and} a lower bound on the average number of transmissions saved via ordering {is} provided. Section \ref{numericalresultssection} presents some numerical results to demonstrate the communication efficiency of the proposed algorithm. Finally, we conclude the paper in Section \ref{conclustionsection}.



\section{Decomposable Graphical Models}\label{DGMsection}

{DGMs have received extensive study in machine learning \cite{xiang1997microscopic,mohan2014node}, sensor networks \cite{cetin2006distributed} and electric power systems \cite{weng2013graphical}}. In this section, we briefly describe {the basic theory of} DGMs.
Consider an undirected graph ${\cal G} = \left( {{\cal V},{\cal E}} \right)$ with $M$ vertices, where ${\cal V} = \{1,2,...,M\}$ is the set of vertices and ${\cal E} =\{(i_1,j_1),(i_2,j_2),...,(i_{|{\cal E}|},j_{|{\cal E}|})\}$ denotes the set of undirected edges of the graph. The graphical model for a random vector ${X_{i,[1,M]}}$ at each time slot $i$ with graph $\cal G$ {describes the statistical dependency model such} that for each time slot $i$, ${X_{i,[1,M]}}$ follows a known distribution {that} obeys the pairwise Markov property with respect to $\cal G$. The distributed observation vector ${X_{i,[1,M]}}$ satisfies the pairwise Markov property with respect to $\cal G$ {if,} for any pair $(m,m')$ of non-adjacent vertices, i.e., $(m,m') \notin {\cal E}$, the corresponding pair of elements of ${X_{i,[1,M]}}$, $X_{i,m}$ and $X_{i,m'}$, are conditionally independent when conditioned on the remaining elements. {This can be} expressed as
\begin{align} \label{pariwise_markov_property}
&f\left( {X_{i,m},X_{i,m'}\left| {{{ {X_{i,{\cal V}\backslash \{m,m'\}}} }}} \right.} \right) \!\notag\\
&= \! f\left( {{X_{i,m}}\left| {{{ {X_{i,{\cal V}\backslash \{m,m'\}}} }}} \right.} \right)  f\left( {{X_{i,m'}}\left| {{{ {X_{i,{\cal V}\backslash \{m,m'\}}} }}} \right.} \right)
\end{align}
{where we have used $f(\cdot)$ to denote the corresponding pdfs.}

An undirected graph is decomposable if the graph has the property that every cycle of length larger
than $3$ possesses a chord \cite{lauritzen1996graphical}. Throughout the paper, we concentrate on decomposable undirected graphical models.  Let $K$ denote the number of cliques in the decomposable undirected graph ${\cal G}$. The sequence of cliques of the graph ${\cal G}$ is denoted by $\{{\cal C}_k\}_{k=1}^K$. We denote the corresponding histories $\{{\cal H}_k\}_{k=1}^K$ and separators $\{{\cal S}_k\}_{k=2}^K$ as
\begin{equation} \label{Define_histories}
{{\cal H}_k} = {{\cal C}_1} \cup {{\cal C}_2} \cup  \cdots  \cup {{\cal C}_k}, \; \forall k=1,2,...,K,
\end{equation}
and
\begin{equation} \label{Define_separators}
{{\cal S}_k} = {{\cal H}_{k - 1}} \cap  {{\cal C}_k}, \; \forall k=2,...,K.
\end{equation}
Note that from (\ref{Define_histories}), the $k$-th history ${{\cal H}_k}$ contains all nodes in the first $k$ cliques. The $k$-th separator ${\cal S}_k$ in (\ref{Define_separators}) is the set of the common nodes between ${{\cal H}_{k - 1}}$ and ${{\cal C}_{k}}$.
A set is complete if it induces a fully connected subgraph \cite{lauritzen1996graphical}. For any decomposable undirected graph $\cal G$, the sequence of cliques $\{{\cal C}_k\}_{k=1}^K$ of $\cal G$ is said to be {\emph{perfect}} if the following conditions are satisfied \cite{lauritzen1996graphical}
\begin{enumerate}
	\item the sets ${\cal S}_k$ are complete for all $k=2,3,...,K$;
	\item for all $k>1$, there is a $j<k$ such that ${\cal S}_k \subseteq {\cal C}_j$. \label{running_intersection}
\end{enumerate}
The condition \ref{running_intersection}) is also called the \emph{running intersection property}. Note that ${\cal S}_k$ separates ${{\cal H}_{k - 1}}\backslash{{\cal S}_k}$ and ${{\cal C}_k}\backslash{{\cal S}_k}$ based on (\ref{Define_histories}) and (\ref{Define_separators}) such that all paths from the nodes in ${{\cal H}_{k - 1}}\backslash{{\cal S}_k}$ to the nodes in ${{\cal C}_k}\backslash{{\cal S}_k}$ intersect ${{\cal S}_k}$.

A mapping $q: \{ {2,3,...,K} \} \to \{ {1,2,....,K} \}$ is defined to specify an association between each separator set and one unique clique such that \cite{zhang2017ordering}
\begin{equation} \label{Define_q_k}
q\left( k \right) \buildrel \Delta \over = \min \left\{ {j \left| \; {{{\cal S}_k} \subseteq {{\cal C}_j}} \right.} \right\}, \; \forall k=2,3,...,K.
\end{equation}
Note that $q(k)$ describes the minimum index of a clique that contains the $k$-th separator ${\cal S}_k$.
Thus, the $k$-th separator ${\cal S}_k$ is associated with the $q(k)$-th clique ${{\cal C}_{q(k)}}$ according to
\begin{equation} \label{S_k_C_k}
{{\cal S}_k} \subseteq {{\cal C}_{q(k)}}.
\end{equation}
{The} $k$-th separator ${\cal S}_k$ is not only contained in the $q(k)$-th clique ${\cal C}_{q(k)}$, but {it is} also contained in the $k$-th clique ${{\cal C}_k}$ {based on} (\ref{Define_separators}), that is,
\begin{equation} \label{S_k_C_k1}
{{\cal S}_k} \subseteq {{\cal C}_k}.
\end{equation}
It {follows} that for any $k>1$, $q(k)$ must exist and
 \begin{align}\label{q_k_smaller_k}
 q(k)<k
\end{align}
for any decomposable undirected graph $\cal G$. Let ${\cal Q}_j$ denote the set of indices of the separators which are associated with the $j$-th clique via the mapping $q$ in (\ref{Define_q_k}), that is,
\begin{equation} \label{Define_Qj}
{{\cal Q}_j} \buildrel \Delta \over = \left\{ {k\left| \; {q\left( k \right) = j} \right.} \right\}.
\end{equation}
Note that ${{\cal Q}_j}$ enumerates the separators contained in the $j$-th clique except the $j$-th separator.
Thus, ${{\cal Q}_j}\cup\{j\}$ contains all the indices of the separators which are contained in the $j$-th clique.
From (\ref{q_k_smaller_k}), we know that the minimal element in ${{\cal Q}_j}$ satisfies
\begin{equation}\label{Q_j_larj}
\min {{\cal Q}_j} > j,
\end{equation}
which implies that
\begin{equation} \label{Q_j_subset}
{{\cal Q}_j}  \subseteq \left\{ {j + 1,j + 2,...,K} \right\}, \; \forall j=1,2,...,K-1,
\end{equation}
and
\begin{equation}\label{emptydefin}
{{\cal Q}_K}  = \emptyset.
\end{equation}

\begin{figure}[!t]
\centering
\includegraphics[width=2.5in]{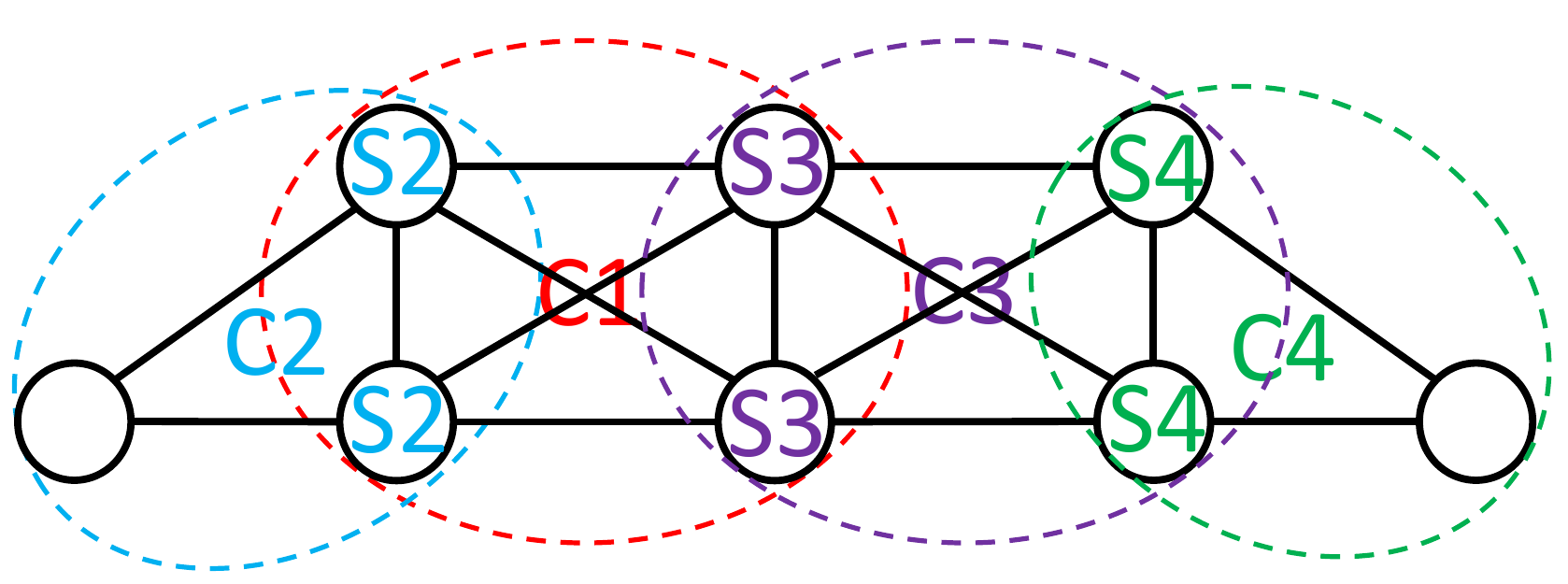}
\caption{The decomposable graphical model with 4 cliques and numbered separators.}
\label{conv_fig0}
\end{figure}

\subsubsection{Illustration of (\ref{Define_histories})--(\ref{emptydefin}) using Fig. \ref{conv_fig0}}

Consider the example in Fig. \ref{conv_fig0}. By employing (\ref{Define_q_k}), we observe that $q(2)=q(3)=1$ and $q(4)=3$ which implies ${{\cal S}_2} \subseteq {{\cal C}_{1}}$, ${{\cal S}_3} \subseteq {{\cal C}_{1}}$ and ${{\cal S}_4} \subseteq {{\cal C}_{3}}$. By employing (\ref{S_k_C_k1}), we also obtain ${{\cal S}_2} \subseteq {{\cal C}_{2}}$, ${{\cal S}_3} \subseteq {{\cal C}_{3}}$ and ${{\cal S}_4} \subseteq {{\cal C}_{4}}$. As per (\ref{q_k_smaller_k}), $q(2)<2$, $q(3)<3$ and $q(4)<4$. By employing (\ref{Define_Qj}), we obtain ${{\cal Q}_1}=\{2, 3\}$ and ${{\cal Q}_3}=\{4\}$ but ${{\cal Q}_2} = {{\cal Q}_4} = \emptyset$.

At each time slot $i$, let ${X}_{{i,{{\cal C}_k}}}$ denote the set of observations in $X_{i,[1,M]}$ that come from the nodes in the $k$-th clique. Let ${X}_{{i,{{\cal S}_k}}}$ denote the observations in $X_{i,[1,M]}$ that come from the nodes in the $k$-th separator set. For any DGM, from the fact that the ordered sequence of cliques ${\cal C}_1, {\cal C}_2,...,{\cal C}_K$ forms a perfect sequence, the joint distribution of $X_{i,[1,M]}$ follows the factorization \cite{lauritzen1996graphical}\footnote{$X_{i,[1,M]}$ following a decomposable graph is only a sufficient condition to guarantee (\ref{decomposablepdf}). In other cases we can attempt to verify (\ref{decomposablepdf}) directly.}
\begin{equation}\label{decomposablepdf}
f\left( X_{i,[1,M]} \right) = \frac{{\prod\limits_{k = 1}^K {f\left( {{{X}_{{i,{{\cal C}_k}}}}} \right)} }}{{\prod\limits_{k = 2}^K {f\left( {X}_{{i,{{\cal S}_k}}} \right)} }}
\end{equation}
where $f(X)$ denotes the marginal pdf of $X$. Note that (\ref{decomposablepdf}) can be derived using the pairwise Markov property in (\ref{pariwise_markov_property}).

\section{Distributed Computation of the Generalized Likelihood Ratio Method in DGM}\label{LIkelihood}
In this section, we begin {by} reviewing the generalized log-likelihood ratio (GLR) procedure in the QCD problem \cite{mei2010efficient}. The {QCD} problem can be modeled as a hypothesis testing problem, given by
\begin{align}\label{general_signal model}
&H_0:\mbox{no change occurs}\nonumber\\
&H_1:\mbox{change occurs at a finite unknown time slot}\ \tau.
\end{align}
Note that when the change occurs, all sensors are assumed to be affected simultaneously as mentioned in \emph{Assumption \ref{assumption3}}.

The GLR {test statistic} up to the current time slot $n$ for (\ref{general_signal model}) is
\begin{align}
&GLR_n=\max\Bigg\{0,\nonumber\\
&\quad \max _{1 \leq n' \leq n}\log\frac{ \prod_{i=1}^{n'-1}  f_{0}\left(X_{i,[1,M]}\right) \prod_{i=n'}^{n} f_{1}\left(X_{i,[1,M]}\right)}{\prod_{i=1}^{n} f_{0}\left(X_{i,[1,M]}\right)}\Bigg\}\\
&=\max\Bigg\{0,\ \max_{1 \leq n' \leq n}\sum_{i=n'}^n\log\frac{f_1(X_{i,[1,M]})}{f_0(X_{i,[1,M]})}\Bigg\}\label{GLLRproced}
\end{align}
{where $GLR_n=0$ implies the decision maker does not declare change up to the current time slot $n$ and will continue acquiring more observations}. In the optimum centralized QCD approach, at each time slot $n$, each sensor {sends} its observation to the FC. After receiving the
data from all sensors, the FC calculates (\ref{GLLRproced}) and compares it to a threshold to decide whether to {declare a change} or continue {to collect observations}. In particular, the GLR procedure will raise an alarm {at the time given by} {\cite{mei2010efficient}}
\begin{align}\label{centralizedCUSUM}
T_{GLR}(b) = \inf\left\{n\ge1:GLR_n\ge b\right\}
\end{align}
where the constant $b>0$ needs to be chosen properly to satisfy the false alarm constraint in (\ref{falsealarmrate}). The procedure in (\ref{centralizedCUSUM}) is also called the {classical centralized} CUSUM algorithm which is shown to be optimal for (\ref{CADD1}) in \cite{moustakides1986optimal}.

{Next} we introduce our distributed approach. {We} make the following {additional} assumptions throughout the paper.
\begin{assumption} \label{pairwise Markov
property}
At each time slot $i$, we assume that ${X_{i,[1,M]}} \buildrel \Delta \over = {( {{X_{i,1}},{X_{i,2}},...,{X_{i,M}}} )}$ satisfies the {pairwise Markov property} in (\ref{pariwise_markov_property}) with respect to a given decomposable undirected graph ${\cal G} = \left( {{\cal V},{\cal E}} \right)$.
\end{assumption}
\begin{assumption} \label{Assumption_local_communication}
Nodes in the same clique are close so that the energy cost of {intra-clique} communications is negligible compared to that of communications between the cliques to the FC, so we focus on communications between the cliques and the FC.
\end{assumption}
\begin{assumption} \label{set not change}
The sets $\{{\cal S}_k\}_{k=2}^K$ and $\{{\cal C}_k\}_{k=1}^K$ do not change throughout the detection process. {We generalize this in Section \ref{graphchangescase}.}
\end{assumption}

Next we develop our distributed computation approach.
From (\ref{decomposablepdf}), we have
\begin{align}\label{rewritedecompdf}
&\log\frac{f_1(X_{i,[1,M]})}{f_0(X_{i,[1,M]})}\notag\\
&= \log\frac{{\prod\limits_{k = 1}^K {f_1\left( {{X_{i,{{\cal C}_k}}}} \right)} }}{{\prod\limits_{k = 2}^K {f_1\left( {{X_{i,{{\cal S}_k}}}} \right)} }}
\frac{{\prod\limits_{k = 2}^K {f_0\left( {{X_{i,{{\cal S}_k}}}} \right)} }}{{\prod\limits_{k = 1}^K {f_0\left( {{X_{i,{{\cal C}_k}}}} \right)} }}\\
&=\sum_{k=1}^K\log\frac{f_1\left( {{X_{i,{{\cal C}_k}}}} \right)}{f_0\left( {{X_{i,{{\cal C}_k}}}} \right)} - \sum_{k=2}^K\log\frac{f_1\left( {{X_{i,{{\cal S}_k}}}} \right)}{f_0\left( {{X_{i,{{\cal S}_k}}}} \right)}\label{logproductsum}\\
&=\log\frac{f_1\left( {{X_{i,{{\cal C}_1}}}} \right)}{f_0\left( {{X_{i,{{\cal C}_1}}}} \right)}- \sum\limits_{k \in {{\cal Q}_1}}{\beta _k}\log\frac{f_1\left( {{X_{i,{{\cal S}_k}}}} \right)}{f_0\left( {{X_{i,{{\cal S}_k}}}} \right)}\notag\\
&\quad+\sum\limits_{j = 2}^K \Bigg( \log\frac{f_1\left( {{X_{i,{{\cal C}_j}}}} \right)}{f_0\left( {{X_{i,{{\cal C}_j}}}} \right)} - {\alpha _j}\log\frac{f_1\left( {{X_{i,{{\cal S}_j}}}} \right)}{f_0\left( {{X_{i,{{\cal S}_j}}}} \right)} \notag\\
&\qquad\qquad- \sum\limits_{k \in {{\cal Q}_j}} {\beta _k}\log\frac{f_1\left( {{X_{i,{{\cal S}_k}}}} \right)}{f_0\left( {{X_{i,{{\cal S}_k}}}} \right)} \Bigg)\label{for_J_k}\\
&=\sum\limits_{k = 1}^K {{L_k}\left( {{X_{{i,{\cal C}_k}}}} \right)}\label{sum_LLLR_local}
\end{align}
where ${\cal Q}_j$ is defined in (\ref{Define_Qj}), and the set of non-negative coefficient pairs $\{(\alpha_k, \beta_k)\}_{k=2}^K$ satisfies
\begin{equation} \label{alpha_beta_1}
{\alpha _k} + {\beta _k} = 1, \; \forall k=2,3,...,K.
\end{equation}
{Note that (\ref{sum_LLLR_local}) expresses} $\log{f_1(X_{i,[1,M]})}/{f_0(X_{i,[1,M]})}$ as a sum of the {clique statistics} ${{L_k}\left( {{X_{{i,{\cal C}_k}}}} \right)}$ for $k=1,2,...,K$ {that} are computed at each clique. {After (\ref{decomposablepdf}) is used to obtain (\ref{rewritedecompdf})}, we can group the separator terms into the associated clique terms based on the results in (\ref{S_k_C_k}) and (\ref{S_k_C_k1}). {In fact}, each separator is a member of several cliques. For any term {involving data} coming from the $k$-th separator set, we can {allocate} $\alpha_k$ percentage of that term to the $k$-th clique and $\beta_k$ percentage to the other cliques {that} also contain the $k$-th separator set. {This allows us to obtain (\ref{for_J_k}) from (\ref{logproductsum})}. The {centralized change detection} test statistic can always be expressed as the sum in (\ref{sum_LLLR_local}) as long as (\ref{alpha_beta_1}) is satisfied. {From (\ref{alpha_beta_1})}, there are uncountably many choices of $\{\alpha_k,\beta_k\}_{k=2}^K$ which {introduces} flexibility in the definition of  ${{L_k}({{X_{{i,{\cal C}_k}}}})}$ {in (\ref{sum_LLLR_local})} while still {ensuring} local computation.
In (\ref{sum_LLLR_local}), ${{L_k}( {{X_{{i,{\cal C}_k}}}})}$  is defined as
\begin{align}\label{statistick1}
{{L_1}( {{X_{{i,{\cal C}_1}}}})}\buildrel \Delta \over = \log\frac{f_1\left( {{X_{i,{{\cal C}_1}}}} \right)}{f_0\left( {{X_{i,{{\cal C}_1}}}} \right)}- \sum\limits_{k \in {{\cal Q}_1}}{\beta _k}\log\frac{f_1\left( {{X_{i,{{\cal S}_k}}}} \right)}{f_0\left( {{X_{i,{{\cal S}_k}}}} \right)}
\end{align}
and for all $j = 2,3,...,K$,
\begin{align}\label{statisticlargerk}
{{L_j}( {{X_{{i,{\cal C}_j}}}})} &  \buildrel \Delta \over = \log\frac{f_1\left( {{X_{i,{{\cal C}_j}}}} \right)}{f_0\left( {{X_{i,{{\cal C}_j}}}} \right)} - {\alpha _j}\log\frac{f_1\left( {{X_{i,{{\cal S}_j}}}} \right)}{f_0\left( {{X_{i,{{\cal S}_j}}}} \right)}\notag\\
&\qquad- \sum\limits_{k \in {{\cal Q}_j}} {\beta _k}\log\frac{f_1\left( {{X_{i,{{\cal S}_k}}}} \right)}{f_0\left( {{X_{i,{{\cal S}_k}}}} \right)}.
\end{align}

{Plugging (\ref{sum_LLLR_local}) and (\ref{GLLRproced}) into (\ref{centralizedCUSUM})} implies that the FC declares a {change at time}
\begin{align}\label{DGM-based CUSUM}
T_{{CS}}(b) = \inf\left\{n\ge1:W_n\ge b\right\}
\end{align}
where the CUSUM statistic $W_n$ is defined as
\begin{align}
W_n\buildrel\Delta\over =
\max \left\{0, \quad \max_{1 \leq n' \leq n}\sum_{i=n'}^n\sum_{k=1}^K{{L_k}\left( {{X_{{i,{\cal C}_k}}}} \right)}\right\}.\label{CUSUMstatisc}
\end{align}
A nice property of the non-negative CUSUM statistic $W_n$ is that it can be computed recursively as
\begin{align}\label{cusumalg}
W_{n}=\max \left\{0, \quad W_{n-1}+\sum_{k=1}^{K} {{L_k}\left( {{X_{{n,{\cal C}_k}}}} \right)}\right\}
\end{align}
with $W_0=0$.
The above recursion is {very useful} {because it requires little memory and is easily updated sequentially}. Instead of directly sending all sensor observations to the FC, the distributed computation method provides the proper way to partition the sensor nodes into $K$ local groups that correspond to the cliques. Each clique $k$ will collect the information from the clique nodes to produce the clique statistic ${{L_k}( {{X_{{n,{\cal C}_k}}}})}$ using (\ref{statistick1}) or (\ref{statisticlargerk}) and then transmit it to the FC. The FC will compute the CUSUM statistic $W_n$ in (\ref{cusumalg}) and {compare} it to the threshold $b$ to decide whether to raise an alarm or continue the process.
We note that when we employ (\ref{DGM-based CUSUM}) with $W_0=0$, $\mbox{WADD}(n')$ in (\ref{WADD}) is equal to \cite{veeravalli2014quickest}
\begin{align}\label{worstcasestartginpoint}
{\mbox{WADD}(n') =  \mathbb{E}_1\left[n'-1\right]}
\end{align}
which {implies that} the worst case detection delay occurs at $\tau=1$. The result in (\ref{worstcasestartginpoint}) makes the computation of the WADD in (\ref{WADD}) straightforward.

While the distributed computation method {takes advantage of the graph structure to aggregate the statistics and avoids unnecessary long range transmissions}, it is also of interest to further reduce the number of {transmissions by the cliques to the FC.  This is addressed in the next section}.

\section{Energy-Efficient QCD using Ordered transmissions}\label{EnergyEfficientQCDOrdering}
In the last section the proposed distributed computation method implements the {optimum centralized} CUSUM algorithm while {taking advantage of the graph structure using cliques}. In order to further {reduce the number of long distance transmissions} {from the cliques to the FC}, in this section we describe an ordered transmission approach, {and} a lower bound on the average number of transmissions saved is provided. The {savings are} shown to be large for cases of interest.

\subsection{Ordered Transmissions for QCD}
The idea of ordered transmissions for QCD is to order and then adaptively halt the transmissions of the clique statistics $\{{{L_k}( {{X_{{n,{\cal C}_k}}}})}\}_{k=1}^K$ {during} each time slot $n$. Specifically, after grouping the nodes into several cliques, the clique with the largest clique statistic magnitude transmits first and the cliques with smaller clique statistic magnitudes possibly transmit later. {This process is repeated during each time slot.} We will show that by sometimes halting transmissions before all $K$ cliques have communicated their clique statistics, {further transmissions} can be saved while achieving the same detection delay as the optimal centralized CUSUM algorithm {that} requires all {nodes} to communicate their {observations} to the FC.

\begin{savenotes}
\begin{algorithm}
\caption{ordered-CUSUM.}\label{COG}
\begin{algorithmic}[1]
\Require{a positive constant $b$.}
\Ensure{$n=0$, $W_0=0$ and a positive number $\eta$.}
\While{$W_n<b$}\footnote{{In practice it may be desired to stop at some large value of time slot n, even if $W_n \ge b$ has not been satisfied}.}
\State The FC updates time slot $n=n+1$ and sets $j=1$.
\For{$k=1,2,...,K$}
\State Clique ${\cal C}_k$ summarizes all the clique sensor data to produce ${{L_k}( {{X_{{n,{\cal C}_k}}}})}$ {as per (\ref{alpha_beta_1})--(\ref{statisticlargerk})} at time $t_n$.
\State Clique ${\cal C}_k$ determines a time $t_{n,k}=t_n+\eta/|{{L_k}( {{X_{{n,{\cal C}_k}}}})}|$ to transmit ${{L_k}( {{X_{{n,{\cal C}_k}}}})}$ to the FC.
\EndFor
\State Order cliques using $t_n< t_{n,k_1}\le t_{n,k_2}\le...\le t_{n,k_K}<t_{n+1}$ {where $k_j$ is the index of the clique which has the $k_j$-th largest $|\hat L_{n,k_j}|$ such that (\ref{orderedLLR}) holds}.
\While{{$j\le K$ }}
\State At time $t_{n,k_j}$, clique $k_j$ transmits $\hat L_{n,k_j}$ to the FC.
\If{$ W_{n,k_j}\le \phi_{n,L}$}
\State The FC decides $W_n=0$.
\State break \textbf{while} loop (line 8).
\Else
\State The FC computes $W_{n,k_j}$ via (\ref{definenewupdatest}).
\EndIf
\State The FC sets $j=j+1$.
\EndWhile
\EndWhile
\State Declare the change occurs at the current time slot $n$ and set $n'=n$.
\end{algorithmic}
\end{algorithm}
\end{savenotes}

Our approach{, which we call ordered-CUSUM,} is summarized in Algorithm \ref{COG}. At the beginning of the current time slot $n$ ({denoted} time $t_n$) each clique $k$ for $k=1,...,K$ determines a time $t_{n,k}=t_n+\eta/|{{L_k}( {{X_{{n,{\cal C}_k}}}})}|$ to transmit {its local statistic} ${{L_k}( {{X_{{n,{\cal C}_k}}}})}$ to the FC, where the positive number $\eta$ can be made as small as the system will allow. Thus {clique transmissions are time ordered using} $t_n< t_{n,k_1}\le t_{n,k_2}\le...\le t_{n,k_K}<t_{n+1}$ where $k_j$ is the index of the clique {which} has the $k_j$-th largest $|\hat L_{n,k_j}|$ {such that}
\begin{align}\label{orderedLLR}
\left|\hat L_{n,1}\right|\ge\left|\hat L_{n,2}\right|\ge...\ge\left|\hat L_{n,K}\right|.
\end{align}
{In this way the cliques with larger-in-magnitude local test statistic transmit earlier}. When the FC receives {a new transmission from a} clique $k_j$, it computes
\begin{align}\label{definenewupdatest}
W_{n,k_j}  \buildrel\Delta\over = W_{n-1} + \sum_{k=1}^{k_j}  \hat L_{n,k}
\end{align}
where $W_{n-1}$ is the CUSUM statistic at time slot $(n-1)$
, {and compares $W_{n,k_j}$ from (\ref{definenewupdatest}) with the updated threshold}
\begin{align}\label{lowerboundthre}
\phi_{n,L} \buildrel\Delta\over= -(K-k_j)\left|\hat L_{n,k_j}\right|.
\end{align}
By sending a message to all cliques, the FC {stops any further clique transmission when} $W_{n,k_j}\le\phi_{n,L}$. {When this occurs the FC declares $W_n = 0$ and the system progresses to the next time slot $t_{n+1}$}. {If all cliques transmit prior to $W_{n,k_j}\le\phi_{n,L}$, then the current time slot is also ended and a decision is made using (\ref{DGM-based CUSUM}) and (\ref{definenewupdatest})}. If all transmission propagation delays are known and timing is synchronized, one can schedule all transmissions back to the FC so they arrive in the correct order. However, {the process can easily be implemented with robustness to small timing errors}.  Even with inaccurate estimates of propagation delays or imperfect  synchronization, since the FC receives the values to be ordered, the FC can put them back in order correctly as long as the FC waits a short period related to the uncertainty. {By design, {the ordered-CUSUM algorithm} will always be optimal, as summarized in the following Theorem.}

\begin{theorem}\label{theorem1}
Consider the QCD problem defined in (\ref{CADD1}). At each time slot $n<\tau$, ordered-CUSUM summarized in Algorithm \ref{COG} achieves the same detection performance as the {optimum centralized approach} while using a smaller average number of transmissions.
\end{theorem}
\begin{IEEEproof}
In the ordered-CUSUM algorithm, during time slot $n$, when the FC receives a new clique {statistic it updates the threshold} $\phi_{n,L}$ via (\ref{lowerboundthre}),
and compares $\phi_{n,L}$ with $W_{n,k'}$ in (\ref{definenewupdatest}).
{Let} the most recent transmission {be given by} $\hat L_{n,k_j}$. {Due to ordering it follows that $|\hat L_{n,k_j}|$} is an upper bound for those of the clique statistics which have not yet been transmitted. {It further follows that} the sum of the clique statistics that have not yet transmitted {must be less than or equal to} $(K-k_j)|\hat L_{n,k_j}|$, which is equal to $-\phi_{n,L}$. If $W_{n,k_j}\le \phi_{n,L}$, then $W_n$ has to be zero according to (\ref{cusumalg}), regardless of the clique statistics that have not yet been transmitted. Hence, even without receiving further transmissions, the FC can implement the {optimum centralized communication unconstrained approach} and declare $W_n=0$ at time slot $n$. On the other hand, the {optimum centralized communication unconstrained approach} continues to transmit at time slot $n$ with nonzero probability. Thus, at each time slot $n$, the average number of transmissions required by ordered-CUSUM is smaller than that of {the optimum centralized communication unconstrained approach} while the detection performance is the same.
\end{IEEEproof}

While we have shown that the ordered-CUSUM algorithm built on ordering is more communication-efficient than the {optimum centralized communication unconstrained approach}, it is interesting to consider whether there exists a lower bound on the average number of transmissions saved by the ordered-CUSUM algorithm. This question will be addressed in the next subsection.

\subsection{Lower Bound on the Average Number of Transmissions Saved}

In this subsection, we derive a
lower bound on the average number of transmissions saved by ordered-CUSUM. The following theorem formally describes the communication saving lower bound for each time slot $n$.

\begin{theorem}\label{generalLowerBoundtheorem}
Consider the QCD problem described in (\ref{CADD1}). When using ordered-CUSUM, for any choice of the pairs $\{(\alpha_k,\beta_k)\}_{k=2}^K$ with $K\ge2$ and $\alpha_k+\beta_k=1$, the average number of transmissions saved $S_n$ for time slot $n$ is bounded from below by
\begin{align}\label{averagesavinginequality}
\!S_{n} \!>\!\left(\left\lceil \frac{K}{2} \right\rceil-1\right)\operatorname{Pr}\left( W_{n-1}=0 \ \mbox{and}\  {{L_k}( {{X_{{n,{\cal C}_k}}}})}<0,\forall k \right).
\end{align}
\end{theorem}
\begin{IEEEproof}
According to ordered-CUSUM, if $W_{n,k_j}$ defined in (\ref{definenewupdatest}) is smaller than the threshold $\phi_{n,L}$ in (\ref{lowerboundthre}), then transmissions will be stopped {during} time slot $n$ and the algorithm proceeds to the next time slot $(n+1)$. Let
\begin{align}\label{firsttimestop}
k_n^{*}\buildrel\Delta\over=&\min\left\{1\le k'< K:  W_{n-1}+\sum_{k=1}^{k'} \hat L_{n,k} \right.\nonumber\\
&\qquad\qquad\qquad\left.<-(K-k')\left| \hat L_{n,k'} \right|\right\}
\end{align}
denote the number of necessary transmissions during time slot $n$ using ordered-CUSUM, {and define}
\begin{align}\label{averagesaving}
S_{n}\buildrel\Delta\over= \mathbb{E}\left[K-k_n^{*}\right]
\end{align}
as the average number of transmissions saved at time slot $n$. Then from (\ref{averagesaving}), $S_{n}$ can be bounded from below by
\begin{align}
S_{n} &= \sum_{k=1}^K(K-k)\operatorname{Pr}(k_n^{*}=k)\label{definiationsav}\\
&\ge \sum_{k=1}^{\left\lfloor K/2\right\rfloor+1} (K-k)\operatorname{Pr}(k_n^{*}=k)\label{dropnonegat}\\
&>\left(\left\lceil \frac{K}{2} \right\rceil-1\right)\sum_{k=1}^{\left\lfloor K/2\right\rfloor+1}\operatorname{Pr}(k_n^{*}=k)\label{boundcoeff}\\
&=\left(\left\lceil \frac{K}{2} \right\rceil-1\right)\operatorname{Pr}\left(k_n^{*} \le \left\lfloor \frac{K}{2}\right\rfloor+1\right).\label{averagesavinginequalityAA}
\end{align}
The result in (\ref{dropnonegat}) is obtained by dropping some non-negative terms in (\ref{definiationsav}). In going from (\ref{dropnonegat}) to (\ref{boundcoeff}), we bound $(K-k)$ by using $(\lceil {K}/{2}\rceil-1)$ for $k=1,...,\lfloor K/2\rfloor+1$.
{Plugging the definition of $k_n^{*}$ {from} (\ref{firsttimestop}) into (\ref{averagesavinginequalityAA})}, we obtain
\begin{align}
&S_{n} >\left(\left\lceil \frac{K}{2} \right\rceil-1\right)\operatorname{Pr}\Bigg( W_{n-1}+ \sum_{k=1}^{\left\lfloor K/2\right\rfloor+1} \hat L_{n,k} \nonumber\\
&\qquad\qquad\le -\Big(\left\lceil \frac{K}{2} \right\rceil-1\Big)\left| \hat L_{n,{\left\lfloor \frac{K}{2}\right\rfloor+1}} \right|\Bigg)\label{Savingstep0}\\
&\ge \left(\left\lceil \frac{K}{2} \right\rceil-1\right)\operatorname{Pr}\Bigg( W_{n-1}=0,\ \hat L_{n,k}<0,\forall k, \mbox{and}\nonumber\\
&\quad  W_{n-1}+ \sum_{k=1}^{\left\lfloor K/2\right\rfloor+1} \hat L_{n,k}\le -\Big(\left\lceil \frac{K}{2} \right\rceil-1\Big)\left| \hat L_{n,{\left\lfloor \frac{K}{2}\right\rfloor+1}} \right|\Bigg)\label{Savingstep1}\\
&\ge\left(\left\lceil \frac{K}{2} \right\rceil-1\right)\operatorname{Pr}\left( W_{n-1}=0 \ \mbox{and}\ \hat L_{n,k}<0,\forall k \right)\label{Savingstep2}\\
&={\left(\left\lceil \frac{K}{2} \right\rceil-1\right)\operatorname{Pr}\left( W_{n-1}=0 \ \mbox{and}\ {{L_k}( {{X_{{n,{\cal C}_k}}}})}<0,\forall k \right)}\label{SavingstepFinally}.
\end{align}
In going from (\ref{Savingstep0}) to (\ref{Savingstep1}), we add two extra constraints which will maintain or reduce the probability. In  (\ref{Savingstep1}), when $W_{n-1}=0$ and $\hat L_{n,k}<0,\forall k$ are true, $W_{n-1}+ \sum_{k=1}^{\lfloor K/2\rfloor+1} \hat L_{n,k}\le -(\left\lceil {K}/{2} \right\rceil-1)| \hat L_{n,{\left\lfloor {K}/{2}\right\rfloor+1}}|$ must be true which implies the result in (\ref{Savingstep2}). {In going from (\ref{Savingstep2}) to (\ref{SavingstepFinally}), we use the fact that all of the ordered statistics $\hat L_{n,k}$ being negative implies all of the original unordered statistics ${{L_k}( {{X_{{n,{\cal C}_k}}}})}$ are negative.} This completes the proof.
\end{IEEEproof}

We point out that the lower bound in (\ref{SavingstepFinally}) is very general {and} is valid for any DGM and any choice of the set of non-negative coefficient pairs $\{(\alpha_k,\beta_k)\}_{k=2}^K$ {with} $\alpha_k+\beta_k=1$. The result in (\ref{SavingstepFinally}) indicates that the lower bound {on $S_n$} depends on the number of cliques {and the joint statistics of} $W_{n-1}$ and ${{L_k}( {{X_{{n,{\cal C}_k}}}})}$. {In the next subsection we show that the savings can be large for several cases of interest}.

\subsection{Large Saving
Gains for Several Cases of Interest}
{Consider} a distance measure $s$ between the distributions of the sensor observations before and after the change time $\tau$. {The} distance measure $s$ is assumed to satisfy the following mild {condition}.
\begin{assumption}\label{distanceAssumption}
For the hypothesis testing problem considered in (\ref{general_signal model}) {with ${{L_k}\left( {{X_{{n,{\cal C}_k}}}} \right)}$ as per (\ref{alpha_beta_1})--(\ref{statisticlargerk})}, we assume that the probability $\operatorname{Pr}({{L_k}\left( {{X_{{n,{\cal C}_k}}}} \right)}<0)\rightarrow 1$ as $s\rightarrow\infty$ {for all $k=1,...,K$ and $n<\tau$}.
\end{assumption}

Intuitively, with a large distance between {the distributions of} the sensor observations before and after the change time, it should be easy for the FC to decide when the change occurs. At the end of this subsection, we provide two popular general QCD problems and the corresponding distance measure to illustrate that \emph{Assumption \ref{distanceAssumption}} is reasonable.

{Under \emph{Assumption \ref{distanceAssumption}}, the} following theorem describes the limiting behavior of the lower bound on the total number of transmissions saved by ordered-CUSUM.
\begin{theorem}\label{limitingsavgintheorem}
Under \emph{Assumptions \ref{assumption3}--\ref{distanceAssumption}}, consider the approach in Algorithm \ref{COG} for the QCD problem in (\ref{CADD1}). With a sufficiently large $s$, the total number of transmissions saved over the {optimum centralized communication unconstrained approach} increases at least as fast as proportional to $K$ while the detection delay is not affected. In particular, the total number of transmissions saved  is lower bounded by $(\left\lceil K/2\right\rceil-1)(\tau-1)$.
\end{theorem}
\begin{IEEEproof}
From {(\ref{averagesavinginequality}) in} \emph{Theorem \ref{generalLowerBoundtheorem}}, we have
\begin{align}\label{resultTheorem2}
S_{n} &>\left(\left\lceil \frac{K}{2} \right\rceil-1\right)\operatorname{Pr}\left( W_{n-1}=0 \ \mbox{and}\  {{L_k}( {{X_{{n,{\cal C}_k}}}})}<0,\forall k \right)
\end{align}
Next we employ induction to show {that} as $s\rightarrow\infty$, for $n<\tau$, we have
\begin{align}\label{}
\operatorname{Pr}( W_{n-1}=0 \ \mbox{and}\ {{L_k}( {{X_{{n,{\cal C}_k}}}})}<0,\forall k )\rightarrow1.
\end{align}
Throughout this proof, we only consider the time slots before the change occurs which means we focus on the case when $n<\tau$.
Specifically, we set $W_0=0$. For a sufficiently large $s$, \emph{Assumption \ref{distanceAssumption}} implies that all clique statistics are negative for $n<\tau$. Thus, the probability $\operatorname{Pr}( W_{0}=0\ \mbox{and}\ {{L_k}( {{X_{{1,{\cal C}_k}}}})}<0,\forall k)\rightarrow 1$ implies $\operatorname{Pr}(W_{1}=0)\rightarrow 1$. If we assume $\operatorname{Pr}( W_{n-2}=0)\rightarrow 1$ at time slot $(n-1)$, then under \emph{Assumption \ref{distanceAssumption}} we have  $\operatorname{Pr}( W_{n-2}=0\ \mbox{and}\ {{L_k}( {{X_{{n-1,{\cal C}_k}}}})}<0,\forall k)\rightarrow 1$ for $n<\tau$ which implies $\operatorname{Pr}(W_{n-1}=0)\rightarrow 1$. Thus, at time slot $n$, we obtain $\operatorname{Pr}(W_{n-1}=0\ \mbox{and}\ {{L_k}( {{X_{{n,{\cal C}_k}}}})}<0,\forall k)\rightarrow 1$ for $n<\tau$.
From (\ref{resultTheorem2}), for all $n<\tau$, we have
\begin{align}\label{limittheoresed}
S_{n}>  \left\lceil \frac{K}{2} \right\rceil-1.
\end{align}
Finally, it follows that the total number of transmissions saved $K_s$ can be bounded by
\begin{align}\label{limittheore1}
K_s> \left(\left\lceil \frac{K}{2} \right\rceil-1\right)(\tau-1).
\end{align}

\end{IEEEproof}

As illustrated by \emph{Theorem \ref{limitingsavgintheorem}}, the total number of transmissions saved by ordering the communications from the cliques to the FC increases at least as fast as linearly proportional to the number of cliques K while achieving the same detection delay as the {optimum centralized communication unconstrained approach}. {\emph{Theorem \ref{limitingsavgintheorem}}} also states that more transmissions can be saved as the change time {increases}. In the following, we provide {two general problems and the corresponding distance measure}.

\begin{example}\label{meanshift}
Consider detecting {a change in the mean of a sequence of sensor observation vectors following a multivariate Gaussian distribution as}\footnote{${{\mathcal N}}\left( {{\bf{a}},{\bf{A }}} \right)$ {denotes} the multivariate Gaussian distribution with mean $\bf a$ and covariance matrix $\bf A$. }
\begin{equation} \label{Hypothesis_testing_problem}
\begin{array}{l}
{X_{n,[1,M]}} \sim {{\mathcal N}}\left( {{\bf{0}},{\bf{\Sigma }}} \right)\ \mbox{when}\ n<\tau\\
{X_{n,[1,M]}} \sim {{\mathcal N}}\left( {{\boldsymbol{\mu }},{\bf{\Sigma }}} \right)\ \mbox{when}\ n\ge\tau
\end{array},
\end{equation}
where ${\boldsymbol{\mu}} \ne {\bf 0}$ and the known covariance matrix $\bf \Sigma$ is assumed to be positive definite. In this problem, we can choose $s=\min_k\| {\bm{\mu}}_{{\cal C}_k} \|$ as the distance measure where {${\bm{\mu}}_{{\cal C}_k}$ denotes the mean vector of the nodes in the $k$-th clique with $\ell_2$-norm $\| {\bm{\mu}}_{{\cal C}_k} \|$}.
\end{example}

\begin{example}\label{chagnecovarianceexaple}
Consider detecting {a change in the covariance matrix of a sequence of sensor observation vectors following a multivariate Gaussian distribution as}
\begin{equation} \label{teststructureDGGM}
\begin{array}{l}
{X_{n,[1,M]}} \sim {{\mathcal N}}\left( {{\bf{0}},{\bf{I }}} \right)\ \mbox{when}\ n<\tau\\
{X_{n,[1,M]}} \sim {{\mathcal N}}\left( {{\boldsymbol{0 }},{\bf{\Sigma }}} \right)\ \mbox{when}\ n\ge\tau
\end{array},
\end{equation}
where ${\bf{I }}$ is an identity matrix and the known covariance matrix $\bf \Sigma$ is assumed to be positive definite. In this problem, $s=\min_{k}\lambda_{\min,k}$ where $\lambda_{\min,k}$ is the minimum eigenvalue of ${\bf{\Sigma}}_{{{\cal C}_k}}$ which denotes the covariance matrix associated with ${X}_{{{n,{\cal C}_k}}}$ for $n\ge\tau$. {The following theorem shows that \emph{Assumption \ref{distanceAssumption}} is valid for \emph{Examples \ref{meanshift}} and \emph{\ref{chagnecovarianceexaple}}.}
\end{example}

\begin{theorem}\label{}
Consider the problems in \emph{Example \ref{meanshift}} and \emph{Example \ref{chagnecovarianceexaple}} and the ordered transmission approach described in the ordered-CUSUM algorithm which employs (\ref{sum_LLLR_local}) with
\begin{equation} \label{alpha_Theorem2}
{\alpha _k} = 1 - {2^{K - k}}\xi,
\end{equation}
and
\begin{equation}  \label{beta_Theorem2}
{\beta _k} = {2^{K - k}}\xi,
\end{equation}
for all  $k=2,3,...,K$ using any $\xi $ which satisfies
\begin{equation} \label{gamma_Theorem2}
\xi  \in \left( {0,\frac{1}{{{2^{K - 1}} - 1}}} \right).
\end{equation}
For any $k=1,2,...,K$ with $K>2$, with sufficiently large $\min_k\| {\bm{\mu}}_{{\cal C}_k} \|$ for (\ref{Hypothesis_testing_problem}) or sufficiently large $\min_{k}\lambda_{\min,k}$ for (\ref{teststructureDGGM}), we have for all $k=1,2,...,K$ and $n<\tau$,
\begin{align}\label{beforechange}
\operatorname{Pr}({{L_k}\left( {{X_{{n,{\cal C}_k}}}} \right)}<0)\rightarrow 1.
\end{align}

\end{theorem}
\begin{IEEEproof}
The proof of this theorem is omitted, since it follows from the proof in \cite{zhang2017ordering} and \cite{chen2019testing}. Specifically, for the problem in (\ref{Hypothesis_testing_problem}), as $s\rightarrow\infty$ with $s=\min_k\| {\bm{\mu}}_{{\cal C}_k} \|$, the {result} in (\ref{beforechange}) can be obtained by following \emph{Theorem 2} in  \cite{zhang2017ordering}. For the problem in (\ref{teststructureDGGM}), as $s\rightarrow\infty$ with $s=\min_{k}\lambda_{\min,k}$, {then} (\ref{beforechange}) {can be} obtained by following \emph{Theorem 3} in \cite{chen2019testing}.
\end{IEEEproof}

\subsection{{Extensions to the Case Where the Graph Structure Changes.}}\label{graphchangescase}

{In this subsection, we generalize \emph{Assumption \ref{set not change}} and consider the case where the graph is fixed under each hypothesis but the graph structure {of $f_0\left( X_{n,[1,M]}\right)$ {(relevant for $n<\tau$)} is not the same as $f_1\left( X_{n,[1,M]} \right)$ {(relevant for $n\ge\tau$)}}}. Suppose the sensor observations $X_{n,[1,M]}$ obey the pairwise Markov property with respect to a decomposable graph ${\cal G}_1 = \left( {{\cal V}_1,{\cal E}}_1 \right)$ before the change ($n<\tau$) and {a decomposable graph} ${\cal G}_2 = \left( {{\cal V}_2,{\cal E}}_2 \right)$ after the change ($n\ge\tau$) with ${\cal G}_1\ne{\cal G}_2$. In order to implement distributed computation and ordered transmissions for this case, the sequence of cliques $\{{\cal C}_k\}_{k=1}^K$ is derived based on ${\cal G} = {\cal G}_1 \cup {\cal G}_2$ instead of ${\cal G}_1$ or ${\cal G}_2$. Compared to ${\cal G}_1$ and ${\cal G}_2$, the graph structure ${\cal G}$ possibly {increases the size of some cliques, implying} extra node data needs to be collected {in} these larger cliques. However, when we employ the known pdf $f_0\left( X_{n,[1,M]} \right)$ or $f_1\left( X_{n,[1,M]} \right)$ in these computations, we use some of the extra data in computations involving $f_0\left( X_{n,[1,M]}\right)$ and the rest in computations involving $f_1\left( X_{n,[1,M]}\right)$. Thus, the graph ${\cal G}$ allows the computations that either ${\cal G}_1$ or ${\cal G}_2$ require.

\begin{figure}[!t]
\centering
\includegraphics[width=3in]{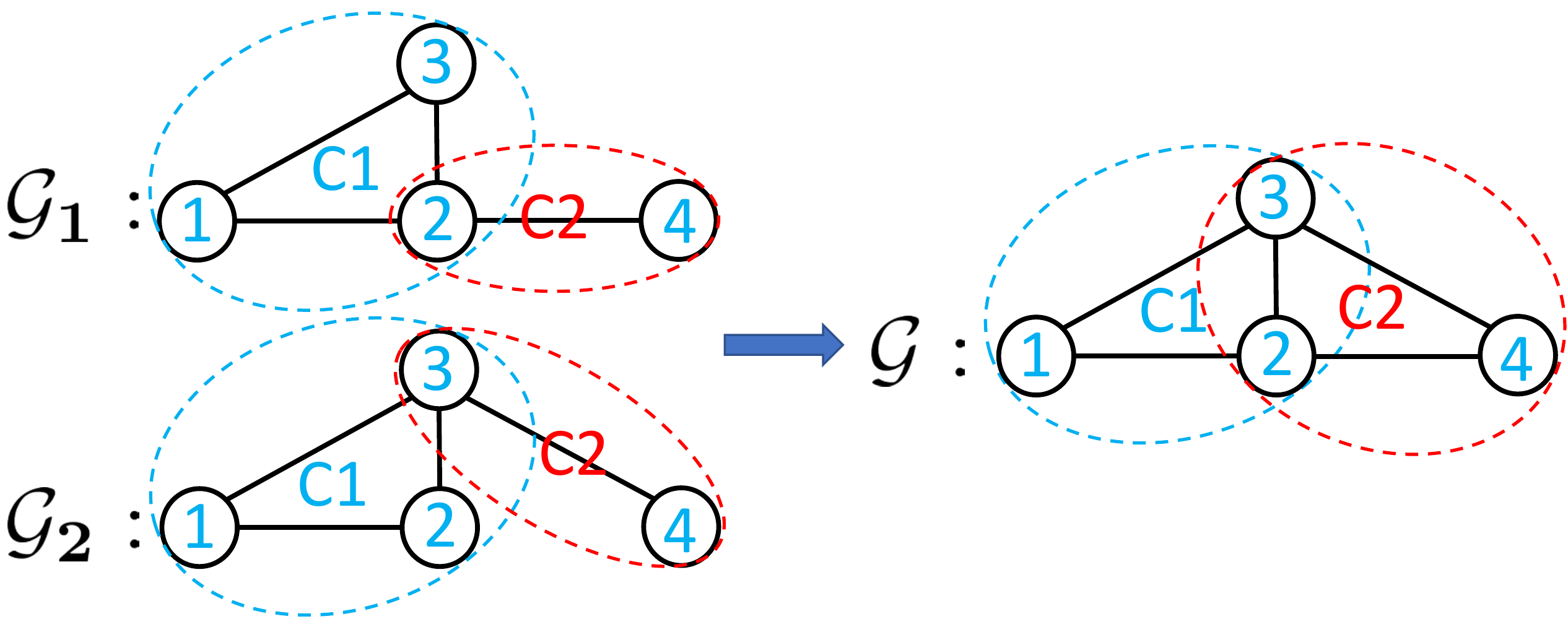}
\caption{{Choice of cliques and separator sets when graph structure changes}.}
\label{conv_fighaha}
\end{figure}

Consider the example illustrated in Fig. \ref{conv_fighaha}. Before the change, the graph structure is indicated by graph ${\cal G}_1$ which has two clique sets ${\cal C}_1=\{1,2,3\}$ and ${\cal C}_2=\{2,4\}$ {along with} one separator set ${\cal S}_2=\{2\}$. After the change occurs, the graph structure illustrated in ${\cal G}_2$ has two clique sets ${\cal C}_1=\{1,2,3\}$ and ${\cal C}_2=\{3,4\}$ along with the separator set ${\cal S}_2=\{3\}$. In order to keep the clique and separator sets the same through the detection process, we collect nodes into {the cliques of} the graph ${\cal G}= {\cal G}_1 \cup {\cal G}_2$ whose clique sets are ${\cal C}_1=\{1,2,3\}$ and ${\cal C}_2=\{2,3,4\}$. {We also employ the} separator set ${\cal S}_2=\{2,3\}$. {These sets} are respectively regarded as the clique sets and the separator set through the whole detection process. Note that compared to ${\cal C}_2=\{2,4\}$ in ${\cal G}_1$, ${\cal G}$ defines a new larger clique ${\cal C}_2=\{2,3,4\}$ to indicate that the data {at} node $3$ {will} be collected {at clique ${\cal C}_2$}. However, when we compute $f_0\left( X_{n,[1,4]} \right)$ in a distributed way, we do not really use the data from node $3$ {in} ${\cal C}_2$ to compute $f_0\left( X_{n,{\cal C}_2} \right)$. Similarly, when we compute $f_1\left( X_{n,[1,4]} \right)$, we do not really use the data from node $2$ {in} ${\cal C}_2$ to compute $f_1\left( X_{n,{\cal C}_2} \right)$.
After implementing the distributed computation {using the {{clique}} and separator sets of ${\cal G}$}, ordered transmissions can be developed according to Section \ref{EnergyEfficientQCDOrdering}. It is worth mentioning that the union operation might {{{decrease}} the number of cliques which can degrade the gains of distributed processing and ordering}.

\section{Numerical Results}\label{numericalresultssection}
In this section, numerical examples for two representative classes of decomposable graphical models (chain structure and tree structure) are presented in order to illustrate the communication saving performance using the proposed ordered transmission approach. Chain structure and tree structure graphs
have been employed in studies on feature representation \cite{liu2010moreau}, topology identification \cite{weng2016distributed}, structure learning \cite{tan2010learning} and electrical power systems \cite{weng2013graphical}.

\subsection{Total number of Transmissions Saved versus the Distance Measure}

\begin{figure}[!t]
\centering
\includegraphics[width=3.5in]{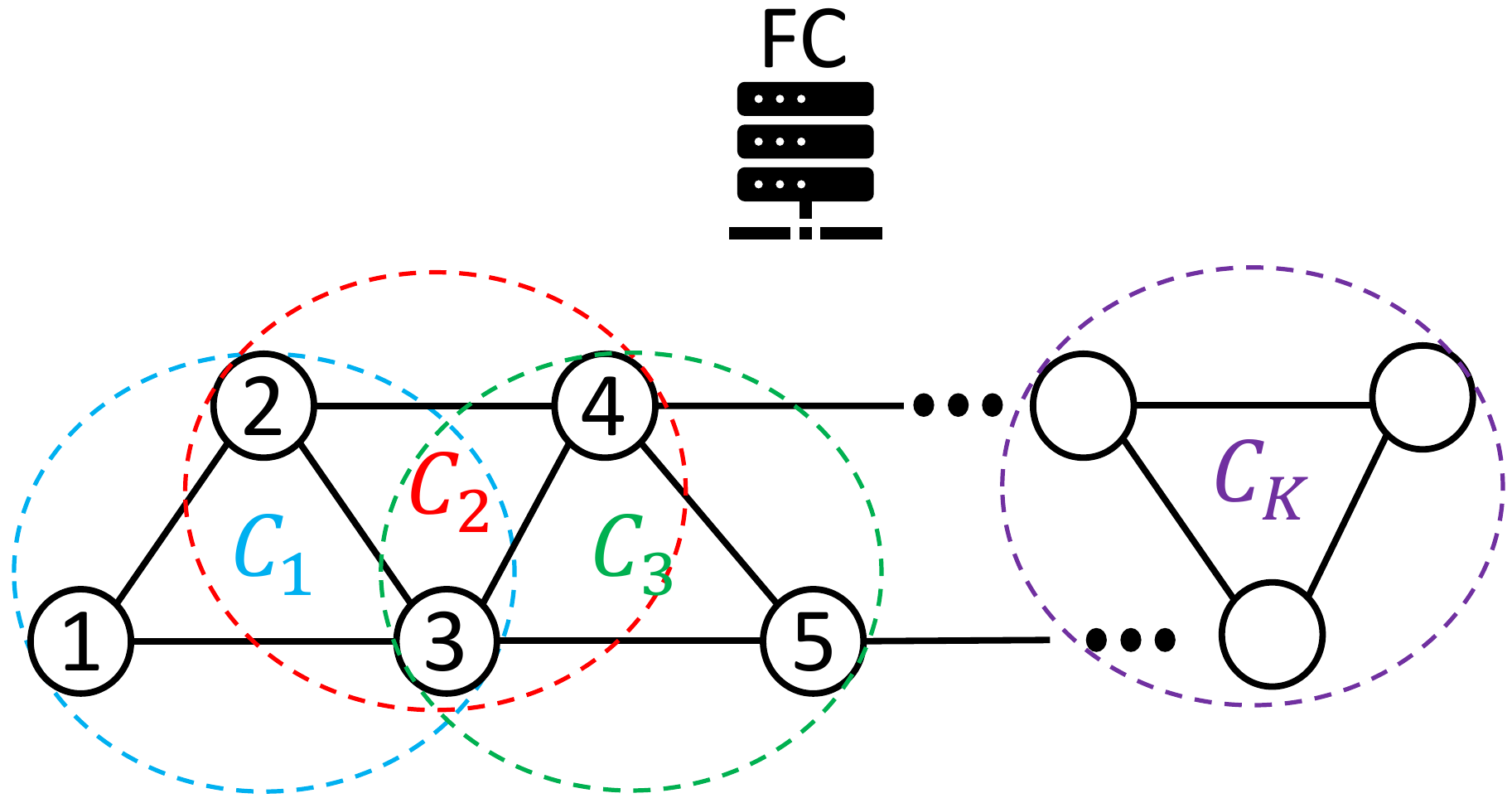}
\caption{The decomposable graphical model with chain structure. Clique 1: Nodes 1,2,3; Separator: Nodes 2,3; Clique 2: Nodes 2,3,4; Separator: Nodes 3,4; and so on down the chain.  Each clique communicates with the fusion center (FC).}
\label{conv_fig1}
\end{figure}

In this subsection, the lower bound in (\ref{averagesavinginequality}) is compared with the actual number of transmissions saved by ordered-CUSUM from Monte Carlo simulations (1000 runs). Consider a {graph} with chain structure as illustrated in Fig. \ref{conv_fig1} where we set the number of cliques $K=50$. As indicated in Fig. \ref{conv_fig1}, each clique has 3 nodes, and every {two-connected clique pair} are coupled through a 2-sensor separator {set}. We first consider {the change detection problem} in (\ref{Hypothesis_testing_problem}) and generate a covariance matrix which satisfies the conditional independence specified by the {graph structure} in Fig. \ref{conv_fig1}. In the simulation results of Fig. \ref{conv_fig1}, we set $\tau=1$, $\xi=0.5/(2^{49}-1)$ and ${\boldsymbol{\mu }}=c[1,1,...,1]^{\top}$. In order to satisfy the false alarm constraint $\mathbb{E}_{\infty}(n)\ge\gamma=10^3$, the minimum value of the positive constant $b$ is found using grid search with the grid points spaced $0.01$ apart. Note that hereafter in the simulation results, we only count the number of transmissions from the cliques to the FC for the first $10^3$ time slots.  In Fig. \ref{conv_fig1}, we plot the total number of transmissions saved versus $c$ when the change does not occur {during these $10^3$ time slots}. Fig. \ref{conv_fig1} indicates that our theoretical lower bound in (\ref{averagesavinginequality}) is valid and its value {increases} as $c$ increases {which means the distance measure $s$ increases}. As expected from our analysis, Fig. \ref{conv_fig1} shows that the lower bound on the total number of transmissions saved nearly equals $24000$ when $c=40$ which is consistent with \emph{Theorem \ref{limitingsavgintheorem}} since $({\left\lceil {{K}/{2}} \right\rceil  - 1})\times 10^3=24000$ when $K=50$.

\begin{figure}[!t]
\centering
\includegraphics[width=3.5in]{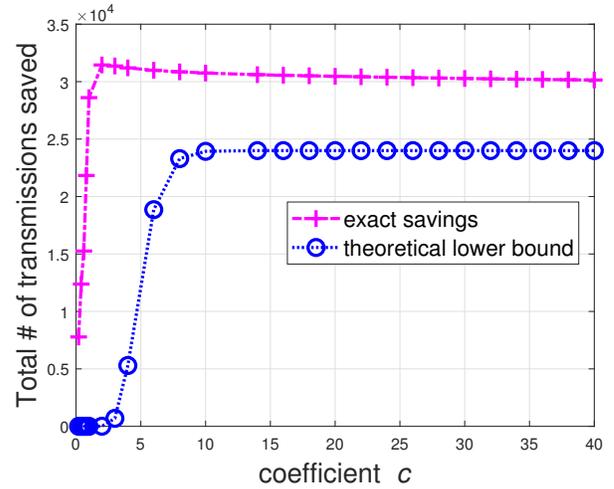}
\caption{Impact of mean shift on the total number of transmissions saved when the change does not
occur {during these $10^3$ time slots}.}
\label{conv_fig2}
\end{figure}

\subsection{Total Number of Transmissions Saved versus the Number of Cliques}

\begin{figure}[!t]
\centering
\includegraphics[width=3.5in]{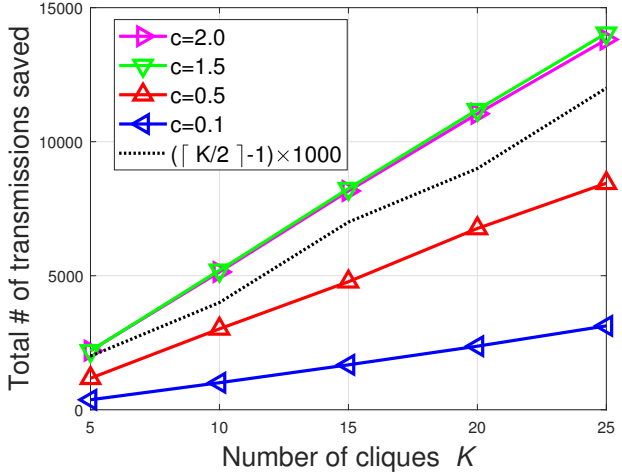}
\caption{The total number of transmissions saved when the change does not occur {during these $10^3$ time slots} versus $K$ for the model
illustrated in Fig. \ref{conv_fig1}.}
\label{conv_fig3}
\end{figure}

{In} this subsection, using Monte Carlo
simulations (1000 runs), we investigate the total number of transmissions saved for the first $10^3$ time slots by ordering the communications from the cliques to the FC for different number of cliques $K$ for the case when no change occurs {during these $10^3$ time slots}. We consider the {testing problem} in (\ref{Hypothesis_testing_problem}) with the same class of {graph structures} as in Fig. \ref{conv_fig1}. We plot the total number of transmissions saved when no change occurs {during these $10^3$ time slots} versus $K$ in Fig. \ref{conv_fig3} for the parameters $\tau=1$, $\gamma=10^3$, $\xi=0.5/(2^{K-1}-1)$ and ${\boldsymbol{\mu }}=c[1,1,...,1]^{\top}$. For comparison, the limiting theoretical lower bound on communication savings in \emph{Theorem \ref{limitingsavgintheorem}} is also provided. For the specific cases considered, Fig. \ref{conv_fig3} indicates that the total number of transmissions saved by Ordered-CUSUM increases approximately linearly with $K$ for every value of $c$. It also indicates that the rate of increase with $K$ increases with increasing $c$ for smaller $c$ {but eventually, the rate of increase saturates as $c$ becomes large, corresponding to a large and easily detectable change}.

\begin{figure}[!t]
\centering
\includegraphics[width=3.5in]{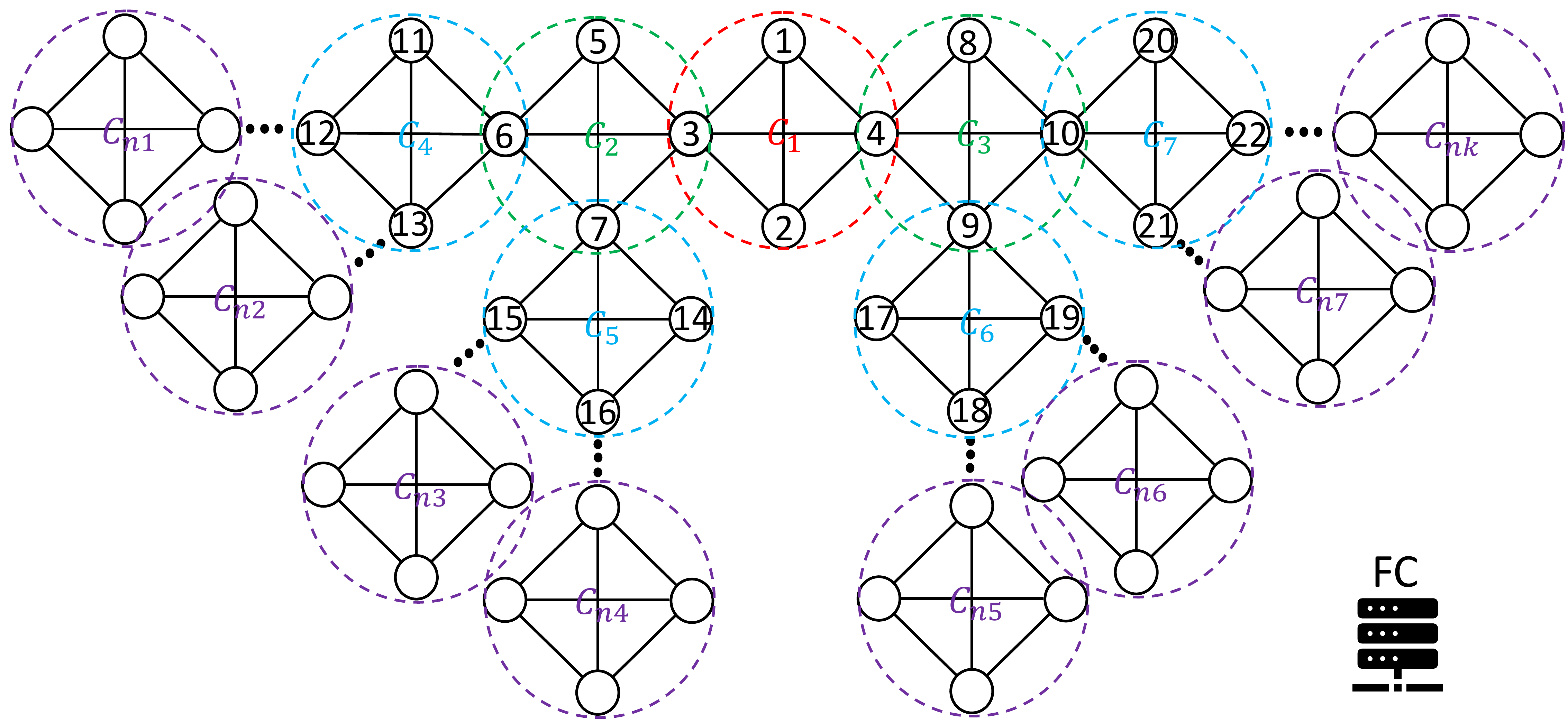}
\caption{The decomposable graphical model with tree structure. Clique 1: Nodes 1,2,3,4; Separator: Nodes 3,4; Clique 2: Nodes 3,5,6,7; Separator: Nodes 6,7; and so on. Each clique communicates with the fusion center (FC).}
\label{conv_fig4}
\end{figure}

\begin{figure}[!t]
\centering
\includegraphics[width=3.5in]{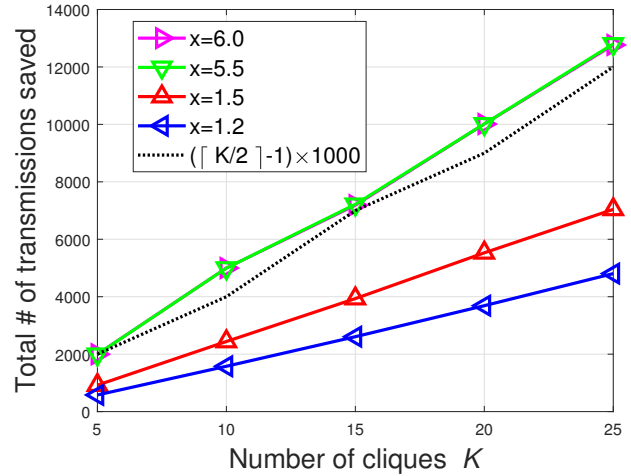}
\caption{The total number of transmissions saved when the change does not occur {during these $10^3$ time slots} versus $K$ for the model
illustrated in Fig. \ref{conv_fig4}.}
\label{conv_fig5}
\end{figure}

Next, we consider a different class of {graphs} with the tree structure as illustrated in Fig. \ref{conv_fig4} where each clique contains 4 nodes and every {two-connected clique pair} are coupled through a 1-sensor separator set. Here we consider the {testing problem} in (\ref{teststructureDGGM}). We set $\tau=1$, $\gamma=10^3$ and $\xi=0.5/(2^{K-1}-1)$.  The diagonal elements of ${\bf\Sigma}_{{\cal C}_k}$ for all $k$ are set to be $x^2$ and the other elements of ${\bf\Sigma}_{{\cal C}_k}$ are set to equal to $x/10$ where {the minimum value of ${\bf\Sigma}_{{\cal C}_k}$ is $x^2-x/10$, so its value may be changed by varying $x$}.  In Fig. \ref{conv_fig5}, we plot the total number of transmissions saved by ordered-CUSUM when the change does not occur {during $10^3$ time slots} versus $K$ for different values of $x$. Fig. \ref{conv_fig5} implies that the total number of transmissions saved increases approximately linearly with $K$ for every value of $x$. Fig. \ref{conv_fig5} also indicates that when $x$ is relatively small {then increasing $x$ increases the slope} which is very similar to the result in Fig. \ref{conv_fig3}.

\section{Conclusion}\label{conclustionsection}
In this paper a new class of communication-efficient QCD schemes for sensor networks have been {developed} that {reduce} the number of transmissions without any impact on detection delay when compared to the optimum centralized communication unconstrained {QCD} approach. {It is assumed that} the observations follow a {decomposable graphical model} (DGM), {which is a very broad class of network topologies, and the observations between sensors may be dependent.} For a QCD problem with sensor observations following any DGM, {we} write the {optimum} centralized change detection test statistic as a sum of clique statistics where each clique statistic can be computed {only using local data available} at the corresponding clique. {To complete the computation of the optimum centralized test}, each clique {forwards} {its} clique statistic to the FC.

In order to further improve the communication efficiency, we have applied the ordered transmission approach over the cliques to reduce the number of transmissions from the cliques to the FC without performance loss. In the ordered transmission approach, the cliques with more informative observations transmit their clique statistics to the FC first. {Transmissions are halted after sufficient evidence is accumulated to save transmissions, {and a new round of sensing is initiated}.} Furthermore, a lower bound on the average number of transmissions saved has been provided. When a well-behaved distance measure between {the pdfs of the} sensor observations before and after the change becomes sufficiently large, the lower bound {approaches} approximately half the number of cliques. Extensions to the case where the graph structure changes have been discussed. In order to illustrate our theoretical analysis, two {{popular general}} QCD problems with sensor observations following a multivariate Gaussian distribution have been considered and numerical results have been provided which are consistent with the analytical findings.



%
%

\ifCLASSOPTIONcaptionsoff
  \newpage
\fi



%
\bibliographystyle{IEEEtran}
\bibliography{refs}

%
%

%

%
%
%





\end{document}